\documentclass[preprint, 5p, number, sort&compress, lefttitle]{elsarticle}

\usepackage{multicol}
\usepackage{amsmath}
\usepackage{amssymb}
\usepackage{graphicx}
\usepackage{bm}
\usepackage{hyperref}
\usepackage{float} 
\hypersetup{
    colorlinks=true,
    linkcolor=blue,
    filecolor=blue,
    urlcolor=blue,
    citecolor=blue,
}

\def\rm{\mathrm}

\newcommand{\eq}[1]{Eq.~(\ref{#1})}
\newcommand{\fig}[1]{Fig.~\ref{#1}}

\usepackage{ulem}
\usepackage{xcolor}
\usepackage{braket}
\newcommand{\new}[1]{{\color{black}{#1}}}
\newcommand{\old}[1]{{\color{red}{{}}}}

\journal{\href{http://arxiv.org/abs/2304.01819}{arXiv:2304.01819}, published in 
\href{https://link.aps.org/doi/10.1103/PhysRevD.108.023002}
{Phys.~Rev.~D 108 (2023) 023002} }

\begin{document}


\begin{frontmatter}

\title{Axion-photon conversion of GRB221009A}

\author[aff1]{Luohan Wang}

\author[aff1,aff2,aff3]{Bo-Qiang Ma\corref{cor1}}
\ead{mabq@pku.edu.cn}
\cortext[cor1]{Corresponding author}

\affiliation[aff1]{
    organization = {School of Physics,
    Peking University},
    city = {Beijing 100871},
    country = {China}}

\affiliation[aff2]{
    organization = {Center for High Energy Physics, Peking University},
    city = {Beijing 100871},
    country = {China}}

\affiliation[aff3]{
    organization = {Collaborative Innovation Center of Quantum Matter},
    city= {Beijing},
    country = {China}}

\begin{abstract}
The newly observed gamma ray burst GRB221009A exhibits the existence of 10~TeV-scale photons, and the axion-photon conversion has been suggested as a candidate to explain such energetic features of GRB221009A. In this work we adopt a model to calculate the conversion probability of the energetic photons from GRB221009A to the Earth. The result shows that the penetration probability of photons with energy above $10^1$~TeV can be up to $10^{-2}$ to $ 10^{-4}$ depending on the coupling constant $g_{a\gamma}$ and the axion mass $m_a$, together on the magnetic field parameters of the source galaxy of GRB221009A. 
We show that the parameters of the source galaxy, with the magnetic field handled by a cellular model, contribute a lot of uncertainties to the penetration probability, so we have more freedom to reconcile a variety of axionlike particle parameters from other observations with the Large High Altitude Air Shower Observatory observation. By comparing the results in this article with the data from Large High Altitude Air Shower Observatory, we can obtain more precise constraints on the ranges of these parameters.
\end{abstract}


\end{frontmatter}

\section{Introduction}

During the propagation of very-high-energy (VHE, $E_{\gamma} \geq 0.1~$TeV) photons in the Universe, the photon flux should be significantly attenuated by the interaction of these photons with background photons to 
\old{produce} 
\new{annihilate into}
electron-positron pairs~\cite{Gould:1966pza, Fazio:1970pr, Protheroe:2000hp}. This mechanism is supposed to strongly suppress the VHE spectra from distant sources 
and thus to forbid the detection of VHE photons.
However, with the oscillation of photons into \old{axion-like} \new{axionlike} particles (ALPs), such suppression can be avoided~\cite{axion-like}. 

\old{Axion-like particle (ALP)} \new{ALP} is a sort of generalization
of the axion, the pseudo-Goldstone boson associated with Peccei-Quinn symmetry to solve the \old{strong-CP} \new{strong-$CP$} 
 problem~\cite{Peccei:1977hh}.
ALPs are very light pseudoscalar bosons represented by $a$. Numerous ALP couplings to the standard model particles can be considered, including the case where ALPs couple to two photons. The axions that we consider in this work are characterized by their coupling with two photons.
In the presence of transverse magnetic field, photons can translate into axions, and vice visa~\cite{Sikivie:1983ip, Masso:1995tw}. For VHE photons from extragalactic sources, the photons can transition into photon-ALP mixed flux in the source galaxy and the axion part can propagate unimpeded through the intergalactic medium, while the VHE photons are absorbed by the extragalactic background photons. When passing through the Galaxy magnetic field, the back conversion of axions to photons can then bring back a fraction of VHE photons. Thus, the conclusions from QED are significantly modified in the presence of ALPs. The above scenario was \old{firstly} \new{first} proposed to explain the excess of very-high-energy gamma ray \old{which} \new{that} should undergo attenuation by the extragalactic background photons~\cite{DeAngelis:2007dqd},
later with source \old{B-field} \new{$B$ field} or the intergalactic magnetic fields considered~\cite{DeAngelis:2008sk,Sanchez-Conde:2009exi,Mirizzi:2009aj,Dominguez:2011xy}
and then with the Milky Way magnetic field also included~\cite{referee-ref2}.

On October \old{$9^{\text{th}}$} \new{9}, 2022, an extremely powerful gamma-ray\old{-burst} \new{burst} named GRB221009A, located at RA=288.282 and Dec=19.495 with z=0.1505, 
was detected by several observatories. The Large High Altitude Air Shower Observatory (LHAASO) detected 
64000
of very high energy photons above $0.2~$TeV, with the maximum of energy from above 10~TeV up 
to 18~TeV~ \cite{gcn32677,LHAASO-grb221009a-1}. Assuming that the photons observed by LHAASO indeed came from GRB221009A, the detection of such energetic photons is forbidden under standard models~\cite{Li:2022wxc,lihao}, but is reasonable under the hypothesis of the photon-ALP oscillation~\cite{axion-like,Galanti:2022pbg,Baktash:2022gnf,space,Nakagawa:2022wwm}. 
Alternative explanations, such as the Lorentz invariance violation of cosmic high-energy photons~\cite{Li:2022wxc,lihao,Finke:2022swf,Zhu:2022usw,Li:2023rgc}, photonic decay of heavy neutrinos from the GRB source~\cite{Cheung:2022luv,Brdar:2022rhc,Huang:2022udc,Guo:2023bpo} or large uncertainty in the energy reconstruction of the observed events~\cite{Li:2022wxc,lihao}, are also proposed to explain the LHAASO observation of above 10~TeV photon events. 

In this paper, we analyze the penetration probability of TeV-scale photons from GRB221009A 
in the presence of ALPs. The result shows that the penetration probability of photons with energy above $10^1$ TeV can be up to $10^{-2}$ to $10^{-4}$, 
indicating the viability for using
the hypothetical axions
to explain the energetic features of
GRB221009A. What is more, this \old{article} \new{paper} gives the quantitative effects of the source galaxy magnetic field and the choice of 
the axion-photon coupling constant $g_{a\gamma}$
and the axion mass $m_a$ 
on photons
with different energy, and calculates the penetration probability for different parameters. For $g_{a\gamma}\in[0.5, 2.1]\times10^{-11}\ \rm{GeV}^{-1}$ and $m_a\in [0.01, 20]\times10^{-8}\ \rm{eV}$, the maximum of total penetration probability for 18 TeV photons is \old{$10^{-2}$, while the minimum is around $10^{-4}$}\new{$10^{-2}$-$10^{-4}$}, based on the magnetic field parameters chosen at the source galaxy with a cellular model. The maximum saturation back conversion probability in the Milky Way is $3.6\times10^{-2}$, for $g_{a\gamma}$ around $2\times10^{-11}\ \rm{GeV}^{-1}$. By comparing the result in this article to the data from LHAASO, we can give more precise constraints on the ranges of the coupling constant $g_{a\gamma}$ and the mass of axions $m_a$, together on the magnetic field parameters of the source galaxy of GRB221009A.

\section{The propagation of high-energy photons under the Standard Model}
\subsection{Absorption of VHE photons by background photons}
The cosmic microwave background (CMB) lights and extragalactic background lights (EBL) are two significant background radiations in the Universe.
According to the standard model, there exists an interaction that creates an electron-positron pair from two photons. Suppose that a high-energy photon collides with a low-energy background photon and produces an electron-positron pair $\gamma\gamma\rightarrow e^+e^-$. Setting $c=\hslash=k_B=1$ in the natural unit, we have
\begin{equation}
    p^{\mu}p_{\mu} \geq 4m_e^2,
\end{equation}
where $p^{\mu}=p^{\mu}_{\gamma}+p^{\mu}_{\rm{bg}}$, with $p^{\mu}_{\gamma}$ and $p^{\mu}_{\rm{bg}}$ being the four-momentum of the gamma ray photon and the background photon. Therefore, the threshold energy of the reaction is
\begin{equation}
    E_{\rm{thr}}=\frac{2m_e^2}{\epsilon(1-\cos\theta)},
\end{equation}
where $\epsilon$ is the energy of background photons and $\theta$
 is the colliding angle between the three momentum of two photons.

The cross section for the reaction of $\gamma\gamma\rightarrow e^+e^-$ is~\cite{Breit:1934zz}
\begin{equation}
    \sigma_{\gamma\gamma}(v)=\frac{\pi}{2}\frac{\alpha^2}{m_e^2}(1-v^2)\left[(3-v^4)\ln\frac{1+v}{1-v}-2v(2-v^2)\right],
\end{equation}
where $v(E,\epsilon, \theta)=\sqrt{1-\frac{2m_e^2}{E\epsilon(1-\cos\theta)}}$ is the speed of generated electron-positron pair in the center-of-momentum frame.

The optical depth for EBL absorption is
\begin{equation}
\begin{split}
    \tau_{\gamma}(E_\gamma)=\int_0^{z_0}\frac{\text{d}x}{\text{d}z}\text{d}z\int_{-1}^1\frac{1-\cos\theta}{2}\text{d}\cos\theta
    \\    \int_{\epsilon_{\rm{thr}}(E_\gamma(1+z),\theta)}^{\infty}\text{d}\epsilon\ \sigma_{\gamma\gamma}(v (E_\gamma(1+z),\epsilon, \theta))\frac{\text{d}n(\epsilon)}{\text{d}\epsilon},
\end{split}
\end{equation}
where $E_\gamma$ is the observed energy of gamma-ray photons, $n(\epsilon)$ is the number density of EBL photons and\\$\epsilon_{\rm{thr}}=
{2m_e^2}/{E_{\gamma}(1+z)}(1-\cos\theta)$ denotes the threshold energy of background photons reacting with gamma photons of observed energy $E_{\gamma}$ at redshift $z$, and
\begin{equation}
    \frac{\text{d}x}{\text{d}z}=\frac{c}{H_0\sqrt{\Omega_m(1+t)^3+\Omega_{\Lambda}}},
\end{equation}
where $H_0=73\ \text{km}\cdot \text{Mpc}^{-1}\cdot \text{s}^{-1}$ is the Hubble constant, $\Omega_m=0.24$ is the energy density and $\Omega_{\Lambda}=1-\Omega_m$ is the dark energy density (in flat cosmology).

Hence, without the assumption of ALPs, the observed energy intensity is $I_{\rm{ob}}(E_\gamma)=I_0(E_0)e^{-\tau_{\gamma}(E_\gamma)}$, where $I_0$ is the emitted spectrum with initial photon energy $E_0=E_\gamma(1+z)$.

\subsection{Interaction with EBL photons}

\begin{figure}[H]\centering
\includegraphics[width=\linewidth]{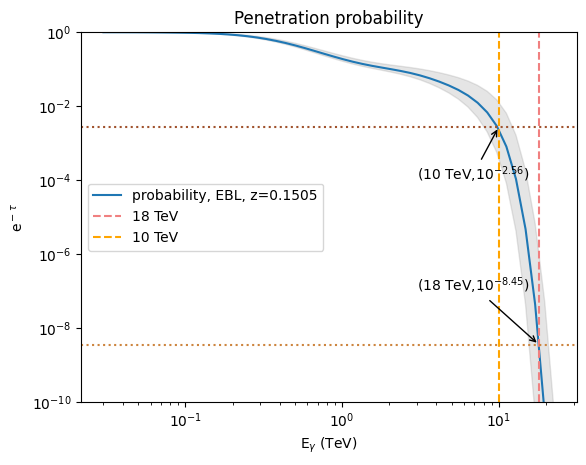}
\caption{The EBL penetration probability $(e^{-\tau_{\gamma}(E_\gamma)})$ as a function of observed gamma-ray energy of TeV photons from $z=0.1505$ under the \old{Standard Model} \new{standard model}. The EBL model is taken from Ref.~\cite{Dominguez2011} and \url{http://side.iaa.es/EBL/}. One sees that 10 TeV scale photons are strongly suppressed by EBL, as already shown in Refs.~\cite{Li:2022wxc,lihao}. }
\label{fig:EBL}
\end{figure}

For CMB photons, the mean energy is  $\epsilon\approx6.35\times10^{-4}$~eV. Therefore, the corresponding threshold energy is $E_{\rm{thr}}^{\text{CMB}}\approx 411$~TeV, which is not in our consideration since the VHE photons detected by LHAASO in GRB221009A have the energy of the order around 10 TeV. EBL photons have the energy between $10^{-3}$ \old{eV} to 1 eV. Consequently, the corresponding threshold energies for collision with EBL photons are between 261 GeV to 261 TeV. Thus, EBL is the dominant contribution to the suppression of 10 TeV-scale photons. The solid 
curve
in \fig{fig:EBL} is the EBL penetration probability $(e^{-\tau_{\gamma}})$ for TeV-scale photons from $z=0.1505$, which is the redshift of GRB221009A. Apparently, VHE photons are strongly suppressed by EBL, and photons with high energy can hardly survive, as already shown in Refs.~\cite{Li:2022wxc,lihao}.

The EBL model picked for the above calculation has been compared with other EBL models in the original EBL model paper in Ref.~\cite{Dominguez2011} and discussed in Refs.~\cite{Li:2022wxc,lihao},
from which we know that the background EBL photons with larger energy are more suppressed than other 
EBL models, so the EBL model we adopted predicts a larger
penetration ability for high energy photons to reach the Earth than other EBL models. Choice of other model calculations would offer more strong conflict between the observation of 18 TeV photons with predictions from \new{the} standard model~\cite{Li:2022wxc,lihao}.

\section{Axions coupling with photons}

In this section we introduce the principle for the $\gamma\gamma a$ coupling process and give the method of calculating conversion probability.

\subsection{General theory}

The axions are characterized by their coupling with two photons, which plays a key role in most searches. The Lagrangian of the interaction is 
\begin{equation}
    \mathcal{L} = -\frac{g_{a\gamma}}{4}F_{\mu\nu}\tilde{F}^{\mu\nu}a = g_{a\gamma}\vec{E}\cdot \vec{B}a,
\end{equation}
where
$F$ is the electromagnetic field-strength tensor and $\tilde{F}_{\mu\nu}=\frac{1}{2}\epsilon_{\mu\nu\rho\sigma}F^{\rho\sigma}$ is its dual, and $g_{a\gamma}$ is the coupling constant of the axion-photon oscillation.

By considering very relativistic axions ($m_a<<\omega$), the equation of the photon and axion amplitudes reduces to a first-order propagation equation. If a beam of polarized monochromatic light with energy $E$ travels along \new{the} $z$-direction, \new{then} we have 
\begin{align}\label{eq:eq1}
    (E-i\partial_z+\mathcal{M})
    \begin{pmatrix}
    A_x\\A_y\\a
    \end{pmatrix}
    =0,
\end{align}
where $E$ is the energy of gamma ray photons, $A_x$ and $A_y$ are the photon polarization amplitudes along \new{the} $x$-axis and \new{the} $y$-axis, $a$ is the ALP amplitude, and $\mathcal{M}$ denotes the photon-axion mixing matrix. The mixing matrix is 
\begin{align}\label{eq:Morigin}
    \mathcal{M}=\begin{pmatrix}
        \Delta_{xx} &\Delta_{xy} &\Delta_{a\gamma}\cos\theta\\
        \Delta_{yx} &\Delta_{yy} &\Delta_{a\gamma}\sin\theta\\
        \Delta_{a\gamma}\cos\theta &\Delta_{a\gamma}\sin\theta&\Delta_a
    \end{pmatrix},
\end{align}
where $\Delta_a=-
{m_a^2}/{2E}$, $\Delta_{a\gamma}=
{g_{a\gamma}B_T}/{2}$, $B_T$ is the transverse magnetic field perpendicular to \new{the} $z$\old{-direction} \new{direction}, and $\theta$ is the angle between $B_T$ and $x$\old{-axis} \new{axis}, i.e. $B_x=B_T\cos\theta, B_y=B_T\cos\theta$. The term $\Delta_{ij}$ mixing $A_x$ and $A_y$ is determined both by the properties of the medium and the QED vacuum polarization effect.

If we set \new{the} $y$\old{-axis} \new{axis} along the direction of $B_T$, \new{then} we can decouple the $x$\old{-component} \new{component}, and \eq{eq:Morigin} reduces to

\begin{align}\label{eq:Mreduce}
    \mathcal{M}=
    \begin{pmatrix}
        \Delta_{\perp}& \Delta_R& 0\\
        \Delta_R & \Delta_{\parallel}& \Delta_{a\gamma}\\
        0& \Delta_{a\gamma}& \Delta_a
    \end{pmatrix},
\end{align}
where $\Delta_{\perp,\parallel}=\Delta_{\rm {pl}}+\Delta^{\rm {CM}}_{\perp,\parallel}$. In a plasma, the photons acquire an effective mass given by $\omega_{\rm{pl}}^2=
{4\pi\alpha n_e}/{m_e}$, leading to $\Delta_{\rm{pl}}=-
{\omega_{\rm{pl}}^2}/{2E}$. $m_e$ is the mass of electron, $n_e$ is the electron density in the medium, and $\alpha$ is the fine structure constant. The $\Delta^{\rm{CM}}_{\perp,\parallel}$ term describes the Cotton-Mouton effect, i.e., the birefringence of fluids in the presence of a transverse magnetic field where $|\Delta^{\rm{CM}}_{\perp}-\Delta^{\rm{CM}}_{\parallel}|\propto B_T^2$. These terms are of little importance to the following topics and can be neglected~\cite{Mirizzi2006}.
$\Delta_R$ is the Faraday rotation term. It depends on the energy and the longitudinal component $B_z$ and causes the coupling between $A_x$ and $A_y$. Since Faraday rotation angle is proportion to $E^{-2}$ and is negligible for VHE photons, $\Delta_R$ can be omitted.

For the relevant parameters, we can numerically calculate
\begin{align}
    \Delta_{a\gamma}&\simeq 3.1\times10^{-2}\left(\frac{g_{a\gamma}}{2\times10^{-11}\ \text{GeV}^{-1}}\right)\left(\frac{B_T}{\rm{\mu G}}\right)\rm{kpc}^{-1}, \\
    \Delta_a &\simeq -7.8\times10^{-3}\left(\frac{m_a}{10^{-8}\ \rm{eV}}\right)^2\left(\frac{E}{\rm{TeV}}\right)^{-1} \rm{kpc}^{-1},\\
    \Delta_{\rm{pl}} &\simeq -1.1\times10^{-10}\left(\frac{n_e}{10^{-3}\ \rm{cm^3}}\right)\left(\frac{E}{\rm{TeV}}\right)^{-1} \rm{kpc}^{-1}.
    \label{eq:parameter}
\end{align}

The coupling constant $g_{a\gamma}\leq 2.1\times 10^{-11}\ \text{GeV}^{-1}$ for $m_a\leq2\times 10^{-7}\ \text{eV}$ is based on the result in Ref.~\cite{Mastrototaro2022}. \old{Ref.} \new{Reference}~~\cite{space}
also gives the condition for $g_{a\gamma}$ and $m_a$ to detect 18 TeV photons. In this \old{article} \new{paper}, we use the combination of conditions from the two articles. 
In the following section, we take 
$g_{a\gamma}=2\times10^{-11}\ \rm{GeV}^{-1}$\old{,} 
\new{and} $m_a=10^{-8}\ \rm{eV}$\old{,} as an approximation for the parameters. 

\subsection{Conversion of $\gamma\gamma\to a$ in the source galaxy} \label{sec3.2}

The gamma ray burst is possibly generated by the collapsing of an extremely massive stellar or a newly born, very rapidly rotating and highly magnetized neutron stars~\cite{Granot2015,MacFadyen1999}. However, in both models, collimated relativistic jets pummel through the surface of its progenitor and radiate as gamma rays. The jet of gamma rays is strongly \old{centred} \new{centered} and propagates through the \old{universe} \new{Universe} as a narrow focused beam. The process of $\gamma\rightarrow a\rightarrow\gamma$ has three steps. \old{Firstly} \new{First}, after going out of the compact region of the source, the photon flux translates into photon-axion mixture due to the interaction with the magnetic filed in the source galaxy. Then the flux propagates through intergalactic area, while VHE photons are absorbed by background photons and the corresponding axions are left undisturbed. Finally, the beam \old{arrive} \new{arrived} at the Milky Way, and the high energy axions can converse back into VHE photons. The analyses for the second and third steps are in \autoref{sec3.3} and \autoref{sec3.4}. In this section, we focus on the conversion of $\gamma\rightarrow a$ in the source galaxy.

Due to the poor knowledge we have about the magnetic field near the GRB source, for simplicity, we can assume that the magnetic field in the source galaxy is composed of many small regions with the same scale $s$. The direction of $\vec{B}$ is the same in each region and changes randomly from one region to another. Such \new{a} model can be called the cellular model (see 
Ref.~\cite{Grossman:2002by}). In fact, we will show that the structure of the cellular model is analogy to our Galaxy magnetic field, so it is reasonable to assume that the source galaxy magnetic field has a cellular structure.

The magnetic field of our Galaxy (i.e., the Milky Way) is relatively well known and can be handled by the Jansson model (see Ref.~\cite{Jansson2012}).
The Galactic magnetic field consists of a regular component and a turbulent component. 
The regular component maintains its strength and direction within several kpc, and the direction of the magnetic field can change abruptly at the interface between different regions, as described in the Jansson model.
From the viewpoint of the cellular model, we can treat the Galactic magnetic field as a composition of many small regions
with the direction and strength of each region magnetic field 
adjusted by the Jansson model. If we only 
concern about an averaged effect such as the propagation of a particle through a number of small regions, we can also adopt the cellular model to treat the magnetic field of each region with random arranged direction and fixed strength.
We will show that  
the regular component of the Milky Way magnetic field can be effectively 
treated by the cellular model with a scale of $s=4$~kpc
to handle the back conversion of axion to photon in
\autoref{sec3.4}.
The turbulent component, with the scale of $10^{-2}$~pc, which dominates on small scale, however, is proved to be irrelevant to the conversion based on the discussion at the end of \autoref{sec3.2}. Therefore, if we assume that the magnetic field of the source galaxy holds similar properties to the Galactic magnetic field, the assumption of a cellular model 
is reasonable.

Therefore, in each small region, we can decouple the $x$\old{-component} \new{component} by setting \new{the} $y$\old{-axis} \new{axis} parallel to the transverse magnetic component $B_T$, and \eq{eq:eq1} reduces to a \old{two dimension} \new{two-dimension} equation consequently,
\begin{align}
        (E-i\partial_z + \mathcal{M}_2)
    \begin{pmatrix}
        A_2 \\ a
    \end{pmatrix},\hfill
\end{align}
with
\begin{align}
    \mathcal{M}_2=
    \begin{pmatrix}
        \Delta_{\rm{pl}}&\Delta_{a\gamma}\\
        \Delta_{a\gamma}&\Delta_a
    \end{pmatrix}.\hfill
\end{align}
We can diagonalize $\mathcal{M}_2$ by a rotation angle 
\begin{equation}
    \theta = \frac{1}{2}\arctan(\frac{2\Delta_{a\gamma}}{\Delta_{\rm{pl}}-\Delta_a}).
\end{equation}
Then we can arrive at the conversion probability ($\gamma$ to $a$) in each small region 
\begin{align}
        P_0(\gamma\rightarrow a)
        &=|\bra{A_2(0)}{a(s)}\rangle|^2,\\
        &=\sin^2(2\theta)\sin^2(\frac{\Delta_{\rm{osc}}s}{2}),\\
        &=(\Delta_{a\gamma}s)^2\frac{\sin^2(\frac{\Delta_{\rm{osc}}s}{2})}{(\frac{\Delta_{\rm{osc}}s}{2})^2},
\end{align}
where the oscillation wave number is
\begin{equation}
    \Delta_{\rm{osc}}^2 = (\Delta_{\rm{pl}}-\Delta_a)^2+4\Delta^2_{a\gamma}.
\end{equation}

If the flux passes through a magnetic field with length $L$ composed of $n$ randomly arranged small regions, \new{then} the total conversion probability is 
\begin{align}
    P_{\gamma\rightarrow a}&=\frac{1}{3}\left[1-(1-\frac{3}{2}P_0)^n\right],\\
    &=\frac{1}{3}\left(1-e^{-\frac{3P_0 n}{2}}\right),\ (P_0\ \mathrm{is \ small}),\label{eq:phost}
\end{align}
the detailed derivation can be seen in Refs.~\cite{Grossman:2002by,Mirizzi2006}. Apparently, $P_{\gamma\rightarrow a}$ reaches its maximum of 1/3 when   $P_0n$ is big enough.

If $2\Delta_{a\gamma}<<|\Delta_{\rm{pl}}-\Delta_a|$, i.e., when $E$ is of GeV scale, $P_0$ is proportional to $(\frac{g_{a\gamma}B_T}{m_a^2})^2\sin^2(\frac{m_a^2}{4E})E^2$, but is relatively small. We can thereby define a critical energy $E_c$ satisfying


\begin{equation}
    |\Delta_a(E = E_c)-\Delta_{\rm{pl}}(E = E_c)| \equiv 2\Delta_{a\gamma},
\end{equation}
below which $P_0$ is negligible. Then $P_0$ can be written as
\begin{equation}
    P_0=(\Delta_{a\gamma}s)^2\frac{\sin^2(\Delta_{a\gamma}s\sqrt{1+(E_c/E)^2})}{(\Delta_{a\gamma}s\sqrt{1+(E_c/E)^2})^2}.
\end{equation}
Note that $E_c\propto m_a^2/g_{a\gamma}B_T$ since $\Delta_{\rm{pl}}$ is negligible compared with $\Delta_{a\gamma}$ and $\Delta_{a}$. For the parameters from \eq{eq:parameter}, and $g_{a\gamma}\sim 10^{-11}\ \rm{GeV}^{-1}$, $m_a\sim 10^{-8}\ \rm{eV}, E_c$ is around 0.1~TeV. Since we expect $\gamma\rightarrow a$ conversion in the source galaxy and back conversion in the Milky Way for VHE photons, $E_c$ is expected to be less than 1 TeV, therefore we have the relation
\begin{equation}
    (\frac{m_a}{10^{-8}\ \rm{eV}})^2(\frac{g_{a\gamma}}{10^{-11}\ \rm{GeV}})^{-1}(\frac{B}{\rm \mu G})^{-1}\leq 
    2\times10^2.
    \label{eq:criticalE0}
\end{equation}

When $2\Delta_{a\gamma}>>|\Delta_{\rm{pl}}-\Delta_a|$ (i.e., $E>>E_c$) or in any case $\Delta_{\rm{osc}}s\rightarrow 0$, the conversion probability $P_0$ is independent of energy, which is just the case in this article. What is more, based on the parameter chosen in this article, the oscillation length $2\pi/\Delta_{\rm{osc}}$ is much larger than the domain size $s$. This means the discontinuity at the interface of two adjacent domains is not felt by the oscillation \cite{Galanti2018}. In this case, the application of the cellular model is successful despite the abrupt change of $B_T$.

\begin{figure}[H]\centering
\includegraphics[width=\linewidth]{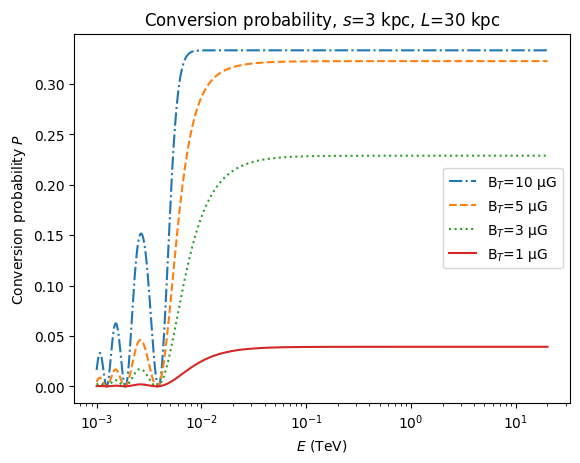}
\caption{The conversion probability of $\gamma$ to $a$ in the source galaxy caused by the regular component with different strength. $B_T \simeq O(\rm{\mu G})$. For photons with energy above 0.1 TeV, $P$ is a constant. For photons with energy below 0.1 TeV, $P$ is an oscillation term dominated by a small coefficient.}
\label{fig:gamma to a}
\end{figure}

The cosmic accelerators producing photons may 
host strong magnetic fields $B\sim O(1\to 10)\ \rm{\mu G}$ (see Refs.~\cite{Galanti2019,Tavecchio2012}).
Suppose that the regular component of magnetic field has $s \simeq 3\ \rm{kpc}$, $L \simeq 30\ \rm{kpc}$, and that $n_e\simeq 1.0 \times 10^{-3} \ \rm{cm^3}$. We can plot the conversion probability of $\gamma$ to $a$ in the source galaxy by the regular component as functions of photon energy $E$ as shown in \fig{fig:gamma to a}.

The conversion probability of $\gamma$ to $a$ for VHE photons is between 0.05 and 0.33 for $B_T$ from $1\ \rm{\mu G}$ to $10\ \rm{\mu G}$. We will take $B_T = 5\ \rm{\mu G}$ as an average field strength \cite{fletcher2011} in the analysis below. We also take $s \simeq 1\ \rm{kpc}$, $L \simeq 10\ \rm{kpc}$ as the minimum typical lengths of the source galaxy, which are actually small parameters for most galaxies.

As to the turbulent component, $s\simeq0.01\ \rm{pc}$, causing $P_0\sim \sin^2(\Delta_{a\gamma}s)\sim(\Delta_{a\gamma}s)^2$ to vanish for the chosen parameters. For example, in the case where $B_T=10\ \rm{\mu G}$, $s=0.01\ \rm {pc}$, $L=30\ \rm {kpc}$, the conversion probability is $10^{-5}$. Therefore, the regular component of magnetic field is the dominant factor in the photon-axion oscillation.

\subsection{Propagation in the intergalactic area}\label{sec3.3}

As has been discussed in Ref.~\cite{axion-like}, the suggestion of adopting the axion-photon conversion as a mechanism to explain the high energy features of GRB 221009A 
requires 
the following conditions: (i) photon-to-axion conversion upon leaving the source galaxy or cluster; (ii) non-significant re-conversion along the intergalactic medium (IG) 
in the path to observer; (iii) axion-to-photon conversion upon entry into the Milky Way.

The intergalactic magnetic field is very weak and has an order of magnitude 
\new{varying from $10^{-11}$ to several}
\old{of} nG.
Assume  
that the strength of magnetic field in intergalactic area is $B_T \sim 1~\rm{nG}$. Therefore, the critical energy $E_c \sim 128$~TeV based on the chosen parameters, is far above the 
detection feasibility. In fact, we expect $E_c$ to be larger than 18~TeV, which is the highest energy for photons detected by LHAASO, and there is 
\begin{equation}
    (\frac{m_a}{10^{-8}\ \rm{eV}})^2(\frac{g_{a\gamma}}{10^{-11}\ \rm{GeV}})^{-1}(\frac{B}{\rm nG})^{-1}\geq 3.5\times10^{-2}.\label{eq:criticalE}
\end{equation}
As a consequence, the photon-axion conversion process
for photons with energy $E<E_c$ is negligible in the intergalactic area. 

Using the combination of \eq{eq:criticalE0} and 
\eq{eq:criticalE}, we can arrive at the same range 
of axion parameters required by the detection of a 18 TeV photon as shown in Fig.~1 in Ref.~\cite{space}. 

During the propagation in the intergalactic area, the intensity of flux is suppressed as shown in \fig{fig:total}. The intensity of the flux arriving at the Galaxy is 
\begin{align}
    &I_{\gamma} = (1-P_{\gamma \rightarrow a})e^{-\tau_\gamma}I_0,\\
    &I_a = P_{\gamma \rightarrow a}I_0,
\end{align}
where $I_0$ is the initial flux intensity at the source.

By ignoring the polarization of photons, the probability amplitude of the mixed flux at the edge of the Milky Way is
\begin{equation}
    \begin{pmatrix}
    A_x\\A_y\\a
    \end{pmatrix}=\begin{pmatrix}
        \sqrt{\frac{I_\gamma}{2}}\\
        \sqrt{\frac{I_\gamma}{2}}\\
        \sqrt{I_a}
    \end{pmatrix}.\label{eq:amplitude}
\end{equation}

\subsection{Back conversion in the Milky Way}\label{sec3.4}

As shown in \autoref{sec3.2}, we can neglect the turbulent component in the Galaxy and discuss the effect of the regular component specifically. 

Measurements from Faraday rotation show that the regular component is basically parallel to the disk, with a small vertical component. By choosing the Jansson model (see Ref.~\cite{Jansson2012}) as the Galaxy magnetic field model, the regular component can be split into a disk part, a toroidal halo part and a X-shaped part. 

The disk component is composed of eight logarithmic spirals located on the Galaxy disk, and the field strength falls off as $r^{-1}$ in each spiral. In the vertical direction, the field strength falls off as a function of $z$, with typical length $h\simeq0.4$~kpc. The electron density inside the Milky Way disk is $n_e\simeq1.1\times 10^{-3}\ \rm{cm^3}$, resulting in $\omega_{\rm{pl}}\simeq 4.1\times10^{-12}\ \rm {eV}$. The toroidal halo component is completely azimuthal. It functions mainly in the area slightly above the disk, and has a typical size of 9.22~kpc ($z>0$) or 16.7~kpc ($z<0$). The X-shaped component has a ``X" shape on the $r$-$z$ plane, and the detailed properties can be seen in Ref.~\cite{Jansson2012}.

In the Galaxy center, there is a possible dipole component with stronger $B$. However, since photons from GRB221009A do not pass through that area, we can focus on the outer regions. What is more, we suppose that the magnetic field strength is zero for $r>20$ kpc.

The propagation of photons in the Milky Way is actually a three dimensional problem, which is more complicated in calculation than simply estimated by the cellular model as we have done in \autoref{sec3.2}.
However, we can still reduce $\mathcal{M}$ to the following expression, which is similar to \eq{eq:Mreduce}

\begin{align}\label{eq:3D}
\mathcal{M}=\begin{pmatrix}
    \Delta_{\rm{pl}} &0 &\frac{g_{a\gamma}}{2}B_x\\
      0 &\Delta_{\rm{pl}} &\frac{g_{a\gamma}}{2}B_y\\
        \frac{g_{a\gamma}}{2}B_x &\frac{g_{a\gamma}}{2}B_y&\Delta_a
        \end{pmatrix}.
\end{align}

Since $E\sim 10^{12} \ \text{eV}>>g_{a\gamma}B\sim 10^{-28}\ \text{eV}$, we can take $\mathcal{M}$ as a perturbation of $E$, $A=A^{(0)}+A^{(1)}$, where $A^{(0)}$ is the zero order term and $A^{(1)}$ is the first order term. The zero order term of \eq{eq:eq1} is
\begin{equation}
    i\partial_z A^{(0)}=EA^{(0)}.
\end{equation}
Therefore, $\new{A^{(0)}(z)}=\mathcal{U}_0(z)A^{(0)}\new{(0)}$, where   $$\mathcal{U}_0(z)=\exp(-i\int_0^z\text{d}z'E).$$

In the interaction representation, we have $A_{\rm{int}}=\mathcal{U}_0^{\dagger}A$, $\mathcal{M}_{\rm{int}}=\mathcal{U}^{\dagger}_0\mathcal{M}\ \mathcal{U}_0=\mathcal{M}$, where $\mathcal{M}$ from \eq{eq:3D} and \eq{eq:eq1} can be written as
\begin{equation}
    i\frac{\partial A_{\rm{int}}}{\partial z} = \mathcal{M}_{\rm{int}}A_{\rm{int}},
\end{equation}
and
\begin{equation}
    A_{\rm{int}}(z) =A_{\rm{int}}(0)-i\int_0^z \text{d}z'\mathcal{M}_{\rm{\rm{int}}}(z')A_{\rm{int}}(z')\label{eq:final},
\end{equation}
where $A^{(0)}_{\rm{int}}(z)=\mathcal{U}_0^{\dagger}A^{(0)}(z)=\mathcal{U}_0^{\dagger}\ \mathcal{U}_0A^{(0)}(0)=A^{(0)}(0)$. 

Preserving the zero order term of $A_{\rm{int}}$, we have $A_{\rm{int}}(0)=A^{(0)}_{{\rm{int}}}(0)=A^{(0)}(0)$. We can numerically calculate \eq{eq:final} based on the Jansson and Farrar model,  with mixing matrix given in \eq{eq:3D} along the galactic line of sight of GRB221009A. The initial probability amplitude is given in \eq{eq:amplitude} and the location of GRB221009A is given in Ref.~\cite{pillera2022}. 

\begin{figure}[H]\centering
\includegraphics[width=\linewidth]{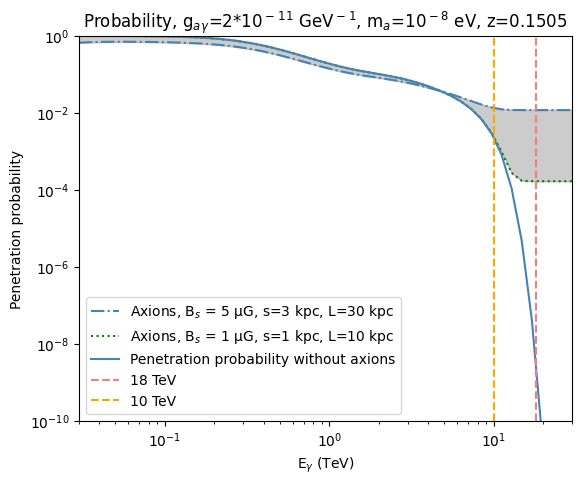}
\caption{The total penetration probability with $g_{a\gamma}=2\times10^{-11}\  \text{GeV}^{-1}, m_a = 10^{-8}\  \text{eV}$ for different parameters at the source galaxy. For $s\leq 1\ \rm{kpc}$, $n={L}/{s}=10$, $B_s\in[1, 10]\ \rm{\mu G}$, the error is shown in the figure. The dot-dashed line is the penetration probability for $B_T=5\ \rm{\mu G}$, $s=3\ \rm{kpc}$, $L=30\ \rm{kpc}$, while the dotted line is the penetration probability for $B_T=1\ \rm{\mu G}$, $s=1\ \rm{kpc}$, $L=10\ \rm{kpc}$.}
\label{fig:total}
\end{figure}

Since there is only a phase shift between $A(z)$ and $A_{\rm{int}}(z)$, the total penetration probability for photons from GRB221009A to the Earth is $|A_{\rm{int},x}(l)|^2 + |A_{\rm{int},y}(l)|^2$, where $l\simeq 23.95$~kpc is the distance from the edge of the Milky Way magnetic field to the Earth. We can plot the total penetration probability as a function of $E_{\gamma}$. For example, assuming that $g_{a\gamma}=2\times10^{-11}\  \text{GeV}^{-1}$, $m_a = 10^{-8}\  \text{eV}$, with varying $B$, $L$, $s$ at the source, we have \fig{fig:total}.

 The upper limit of the probability does not grow with $s>3$~kpc since \eq{eq:phost} has converged to the value of 1/3. For different parameters given in the source, the penetration probability ranges from $10^{-4}$ to $10^{-2}$, depending on the properties of magnetic field at the source galaxy. Apparently, the existence of axion-photon oscillations can strongly enlarge the penetration probability of VHE photons. 
 
\begin{figure}[H]\centering
\includegraphics[width=\linewidth]{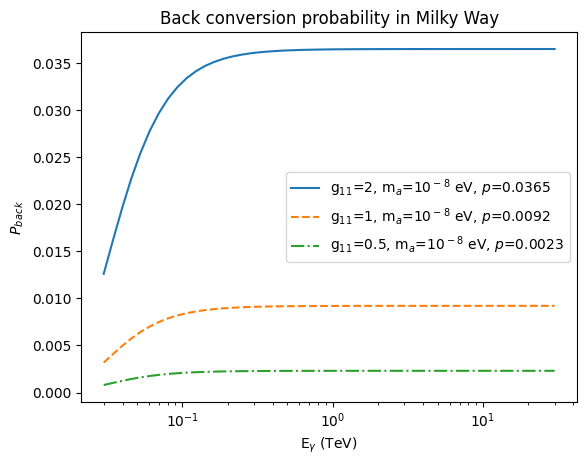}
\caption{The back conversion probability in the Milky Way  with different axion parameters $g_{a\gamma}$ and $m_a$.
In the figure $g_{11}$ denotes $g_{a\gamma}/(10^{-11}~\text{GeV}^{-1})$ and $p$ denotes the corresponding probability at $E_{\gamma}=18~\rm{TeV}$. The maximum of back conversion probability is $3.6\times10^{-2}$.}
\label{fig:MW}
\end{figure}

 The back conversion probability in the Milky Way also converges to a constant as $E_{\gamma}\rightarrow \infty$, which can be seen intuitively in \fig{fig:MW}, where $p$ denotes the back conversion probability at $E_{\gamma}=18~\rm{TeV}$. Defining $g_{11}=g_{a\gamma}/10^{-11}\rm{GeV}^{-1}$, we take $g_{11}\ge 0.5$ so that the back conversion in the Milky Way is not too weak. However, note that the real penetration probability for photons with energy above $E_c$ in the intergalactic magnetic field is not a constant, since these axions may convert into photons and be absorbed by EBL. The correction to the 
 standard model 
 will only be significant for photons with energy between $\rm{max}(E_{c,\rm{source}},E_{c,\rm{MW}})$ and $E_{c, \rm{IG}}$, where $E_{c,\rm{source}}$, $E_{c,\rm{MW}}$, and $E_{c, \rm{IG}}$ denote the critical energy in the source galaxy, the Milky way, and the intergalactic area respectively.

\begin{figure}[H]\centering
\includegraphics[width=\linewidth]{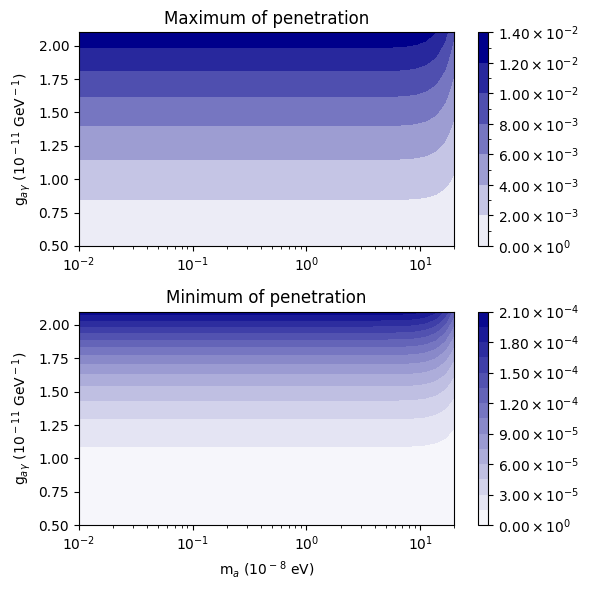}
\caption{The penetration probability for a 18~TeV photon with varying axion parameters $g_{a\gamma}\in [0.5, 2.1]\times10^{-11}\ \rm{GeV}^{-1}$ and $m_a\in [0.01, 20]\times 10^{-8}\ \rm{eV}$. The picture above has $B_T = 10\ \rm{\mu G}$, $L=30\ \text{kpc}$, $s = 3\ \text{kpc}$ corresponding to the maximum of penetration, while the lower picture has $B_T = 1\ \rm{\mu G}$, $L=10\ \text{kpc}$, $s = 1\ \text{kpc}$ corresponding to the minimum of penetration.}
\label{fig:heatmap}
\end{figure}

What is more, we can calculate the penetration probability with varying $g_{a\gamma}$ and $m_a$ for photons with energy of 18 TeV as shown in \fig{fig:heatmap}. We take the values of $g_{a\gamma}$ and $m_a$ from the combined limit of 
Refs.~\cite{Galanti:2022pbg,Baktash:2022gnf,space,Nakagawa:2022wwm} to explain the LHAASO observation and  Refs.~\cite{Eckner2022,Dessert:2022yqq,Dirson:2022lnl} to meet constraints from other astrophysical observations:
$g_{a\gamma}\in[0.5, 2.1]\times10^{-11}\ \rm{GeV}^{-1}$ and $m_a\in [0.01, 20]\times10^{-8}\ \rm{eV}$.
From \fig{fig:heatmap}, we notice that the calculated penetration probability is sensible to the axion-photon coupling $g_{a\gamma}$ but not so sensitive to the axion mass $m_a$.

If we restrict $m_a<2\times10^{-8}\ \rm{eV}$, in the range given in Ref.~\cite{Mastrototaro2022}, \new{then} the penetration probability for 18 TeV photons from GRB221009A is $10^{-2}$ for $g_{11}=2.1$.

We still need the detected data for LHAASO to determine the specific values of $g_{a\gamma}$ and $m_a$, however, the analysis in this article shows that the existence of axions gives the penetration probability of VHE photons within $10^{-2}$ to $10^{-4}$, which does explain the detection of VHE photons, and gives the criterion to judge the ranges of $g_{a\gamma}$ and $m_a$. 

According to \fig{fig:heatmap}, if we know the order of magnitude of the penetration probability, we can basically constrain the parameters at the source, and focus on the choice of $g_{a\gamma}$ and $m_a$. Then we can limit the values of $g_{a\gamma}$ and $m_a$ to more narrow ranges.

The ALP parameters \old{to} \new{can} explain the LHAASO observation \old{vary} \new{variances} with significant differences in literature~\cite{Galanti:2022pbg,Baktash:2022gnf,space,Nakagawa:2022wwm}:  
$g_{a\gamma}\in[0.5, 1]\times10^{-11}\ \rm{GeV}^{-1}$ and $m_a\in [0.01, 20]\times10^{-8}\ \rm{eV}$.
There are also constraints 
from other astrophysical observations, such as these 
from 
the analysis of cosmic sub-PeV gamma rays from the Crab Nebula with $g_{a\gamma}\le 1.8\times10^{-10}\ \rm{GeV}^{-1}$ and $m_a\le 2\times 10^{-7}$--~$6\times 10^{-7}\ \rm{eV}$~\cite{Bi:2020ths},
from galactic
sub-{PeV} gamma rays~\cite{Eckner2022}  with $g_{a\gamma}\le 2.1\times10^{-11}\ \rm{GeV}^{-1}$ for $m_a\le 2\times 10^{-7}\ \rm{eV}$,
from optical polarisation measurements of magnetic white dwarfs~\cite{Dessert:2022yqq} with $g_{a\gamma}\le 5.4\times10^{-12}\ \rm{GeV}^{-1}$ and $m_a\le 3\times 10^{-7}\ \rm{eV}$, and from gamma-ray spectra of flat-spectrum radio
quasars~\cite{Dirson:2022lnl} with $g_{a\gamma}\le 5\times10^{-12}\ \rm{GeV}^{-1}$.
We show that by adjusting the parameters of the source galaxy, we can reproduce the required magnitude to explain the LHAASO observation of above 10 TeV photons with a variety 
of $g_{a\gamma}\in[0.5, 2.1]\times10^{-11}\ \rm{GeV}^{-1}$ and $m_a\in [0.01, 20]\times10^{-8}\ \rm{eV}$, as shown in \fig{fig:MW} and \fig{fig:heatmap}. The novelty of the present work, in comparison with previous literature~\cite{Galanti:2022pbg,Baktash:2022gnf,space,Nakagawa:2022wwm}, reveals the large uncertainties introduced by the parameters of the source galaxy to reconcile a variety of 
axion parameters with the LHAASO observation. 
The parameters of the cellular model about the source galaxy can be further constrained to narrower ranges when there will be more information about the source and the magnetic field of the source galaxy.

In our above calculations, the results are obtained by adopting the realistic Jansson model to treat the magnetic field of the Milky Way. If we adopt the cellular model to calculate the back conversion of axion to photon conversion, we find that we need effective parameters $s=4$~kpc, $B_T=1\ \rm{\mu G}$, $L=20$~kpc for getting the same results obtained with the Jansson model. This proves the efficiency of adopting the cellular model to handle the magnetic field of the Milky Way. Therefore we can use the simple cellular model of galactic magnetic field to handle both the source galaxy and our local Galaxy flexibly. 

\section{Conclusions}

In this work, we use \old{axion-like} \new{axionlike} particles,
which appear in many extensions of the standard model, to explain the energetic features of the newly detected GRB221009A. In the presence of the axion-photon conversion process $\gamma\rightarrow
a\rightarrow \gamma$, we calculate the conversion probability, back conversion probability and the total penetration probability of \old{very-high-energy (VHE)} \new{VHE} photons with varying parameters at the source galaxy, the coupling constant $g_{g\gamma}$, and the mass of axions $g_{a\gamma}$. The presence of axions can significantly increase the penetration probability of VHE photons, with a maximum of $10^{-2}$. Even with the smallest conversion probability assumed in the source galaxy, the penetration probability can reach $10^{-4}$.

Though we have given the penetration probability for different $g_{a\gamma}$ and $m_a$, however, as shown in \fig{fig:heatmap}, the parameters in the source galaxy,
with the magnetic field handled by a cellular model, contribute a lot of uncertainties to the penetration probability. This provides us more freedom to reconcile a variety of \old{axion-like} \new{axionlike} paricle parameters from other observations with the LHAASO observation of above 10 TeV photons.
More precise ranges of the parameters can be obtained by confronting our theoretical prediction with
the data from LHAASO.
Further studies should be based on more experimental results, especially VHE photons from galaxy clusters or other extragalactic sources, in order to reach more precise constraints on $g_{a\gamma}$ and $m_a$, together on the parameter space of the source galaxy.
Alternative possibilities, such as the Lorentz invariance violation or heavy neutrino decay, should be also checked by confronting various features of experimental observations with theoretical predictions comprehensively. 


\section*{Acknowledgments}
This work is supported by National Natural Science Foundation of China (Grant No.~12075003).



\def\urlprefix{}
\def\url#1{}

\end{document}